\documentclass[runningheads,citeauthoryear]{apinv}
\usepackage{epsfig,cite,graphics}

\usepackage[T2A]{fontenc}
\usepackage[cp1251]{inputenc}
\usepackage[bulgarian,english]{babel}

\begin{document}

\title{The new FUor star HBC 722 - one year after the outburst}
\titlerunning{The new FUor star HBC 722 - one year after the outburst}
\author{Evgeni Semkov\inst{1}, Stoyanka Peneva\inst{1}}
\authorrunning{E. Semkov and S. Peneva}
\tocauthor{Evgeni Semkov}
\institute{Institute of Astronomy and National Astronomical Observatory, Bulgarian Academy of Sciences, Sofia, Bulgaria
        \newline
        \email{esemkov@astro.bas.bg, speneva@astro.bas.bg}
}
\papertype{conference talk}
\maketitle

\begin{abstract}
The first results from studies of the new FU Orionis star found in the field of NGC 7000 are presented in the paper. 
During the year passing from the registration of outburst at fourteen papers containing data from observations of this object have been published in the astronomical journals.
These publications present photometric and spectral observations of HBC 722 ranging from the far infrared to X-ray wavelength region.
HBC 722 is the first FU Orionis object, whose outburst was observed from its very beginning in all spectral ranges.
We expect that the interest in this object will increase in the coming years and the results will help to explore the nature of young stars.
\end{abstract}


\Bg
\bgtitle{Новият Фуор HBC 722 - една година след избухването}  
{Евгени Семков, Стоянка Пенева}  
{В статията са представени първите резултати от изследванията на новата звезда от типа FU Orionis открита в областа на NGC 7000.
През изминалата една година след избухването в астрономическите издания са публикувани общо четиринадесет статии с данни от наблюдения на този обект.
Публикациите представят фотометрични и спектрални наблюдения на HBC 722 в диапазона от далечната инфрачервена до рентгеновата област на спектъра. 
HBC 722 е първият обект от типа FU Orionis, чието избухване е наблюдавано от самото му начало във всички спектрални диапазони.
Ние очакваме, че интересът към този обект ще се увеличава през следващите години и неговите изследвания ще помогнат за изучаването на природата на младите звезди.}
\Eng

\section{Introduction}

The study of photometric variability of young stars gives us valuable information about their circumstellar environment.
There are certain evidences from observations that massive circumstellar disks are formed around some of the pre-main sequence (PMS) stars.
The presence of a disk and the interaction between disk and star may be responsible for some properties of PMS stars, as infrared and ultraviolet excesses and stellar wind.
During the final periods of star formation, accretion of material onto the star surface continues partially through the circumstellar disc. 
In the most cases the rate of accretion is slow and does not exceed 10$^{-7}$$M_{\sun}$$/$yr.

One of the most remarkable phenomenon in the early stages of stellar evolution is the FU Orionis (FUor) outbursts.
The flare-up of FU Orionis itself was documented by Wachmann (1939) and for several decades it was the only known object of that type. 
Herbig (1977) defined FUors as a class of young variables after the discovery of outbursts from two new objects - V1057 Cyg and V1515 Cyg. 
The main characteristics of FUors are an increase in optical brightness of about 4-5 mag, a F-G supergiant spectrum with broad blue-shifted Balmer lines, strong infrared excess, connection with reflection nebulae, and location in star-forming regions (Reipurth \& Aspin 2010). 
Typical spectroscopic properties of FUors include a gradual change in the spectrum from earlier to later spectral type from the blue to the infrared, a strong Li I ($\lambda$ 6707) line, P Cygni profiles of H$\alpha$ and Na I ($\lambda$ 5890/5896) lines, and the presence of CO bands in the near infrared spectra (Herbig 1977, Bastian \& Mundt 1985). 
The light curves of FUors are varying from one object to another, but in the most cases the rise goes faster than decline in brightness.
The outbursts of FUors lasting several decades or even a century (Peneva et al. 2010).

The widespread explanation of the FUor phenomenon is a sizable increase in the disc accretion rate onto the stellar surface.
The cause of increased accretion appears to be thermal or gravitational instability in the circumstellar disk (Hartmann \& Kenyon 1985, 1996).
During the outburst, accretion rates raze from $\sim$10$^{-7}$$M_{\sun}$$/$yr to $\sim$10$^{-4}$$M_{\sun}$$/$yr which changes significantly the circumstellar environment. 
The surface temperature of the disk becomes 6000-8000 $K$ and it radiates most of its energy in the optical wavelengths.
For the period of $\sim$100 years the circumstellar disk adds $\sim$10$^{-2}M_{\sun}$ onto the central star and it ejects $\sim$10\% of the accreting material in a high velocity stellar wind. 

Regardless of the significant interest in study of FUors only a dozen objects are certainly assigned to this class of young variable stars (Reipurth \& Aspin 2010, Kazarovets et al. 2011).
Another ten objects show the same spectroscopic characteristics as FUors, but for which an eruption has not been observed (Reipurth \& Aspin 2010).
Registration of an outburst in the optical wavelengths is considered to be a necessary condition a single PMS star to be accepted as FUor.
Therefore, any new announcement for registration of an eruption from PMS star is welcomed with a great interest by researchers.

\section{The outburst of HBC 722}

During our optical photometric monitoring of star-forming regions in 2010, we discovered a large amplitude outburst from a PMS star located in the dark clouds (so-called "Gulf of Mexico") between NGC 7000 (the North America Nebula) and IC 5070 (the Pelican Nebula). 
The star is a member of a small group of H$\alpha$ emission objects in the vicinity of LkH$\alpha$ 188, a region characterized by active star formation (Fig. 1).
Before the outburst the star was not registered as variable and there are not published data from photometric studies.
The only available spectroscopic study of this object was published by Cohen \& Kuhi (1979), who observed the star in 1977.  
They classified the absorption spectrum as K7-M0 and listed the equivalent widths of the H$\alpha$, H$\beta$, and [OI] 6300 spectral lines, the only three lines visible in emission in their spectra.  
The star is included in the catalogs of emission-line stars compiled by Herbig \& Bell (1988), with designation HBC 722.
Due to the great interest in FUor objects and the expected large number of publications the star received the GCVS name V2493 Cyg even before publication of the regular 80th Name List of Variable Stars (Kazarovets et al. 2011).

\begin{figure}[!htb]
  \begin{center}
    \centering{\epsfig{file=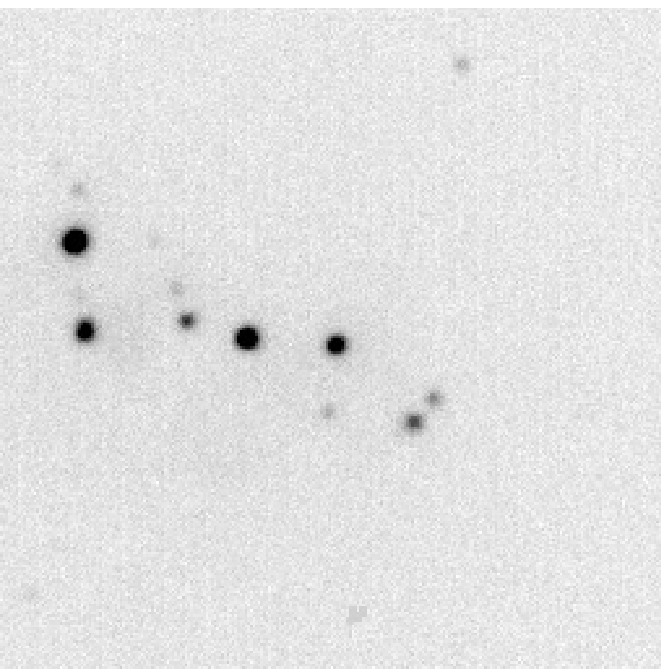, width=0.4\textwidth} \hspace{5 mm} \epsfig{file=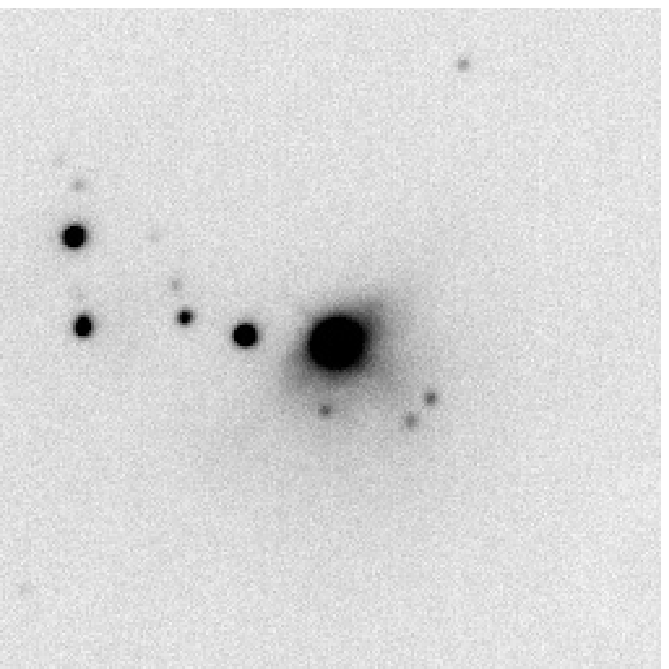, width=0.4\textwidth}}
    \caption[]{CCD frames of HBC 722 obtained with the Skinakas Observatory
   1.3-m RC telescope through a $R$ filter.  {\it Left}: on 2009 Jul. 31. 
   {\it Right}: on 2010 Aug.  26.  The appearance of small reflection nebula
   around the star is observed on the second frame.}
    \label{countryshape}
  \end{center}
\end{figure}

The announcement of the outburst discovery and our first results from photometric observations was published in Semkov \& Peneva (2010a, 2010b).
Fig. 1 compares two $R$-band CCD images obtained before and during the outburst of HBC 722.  
The emergence of a weak reflection nebula around the star is visible in the second image.  
The first spectroscopic observations during the outburst showed that the star's spectrum has changed significantly and it becomes similar to FUor  spectrum in the optical wavelengths (Munari et al. 2010).
Registration of the outburst in the infrared wavelengths on the basis of comparison the observed $JHK$ values with those obtained in 2MASS catalog was reported by Leoni et al. (2010). 
Using the Swift X-ray Telescope Pooley \& Green (2010) detected HBC 722 in the 0.2-10 keV band, the first X-ray detection of a FUor object during the outburst.

\section{Observations of HBC 722 from Rozhen and Skinakas observatories}

Our observations were performed with the telescopes of the National Astronomical Observatory Rozhen (Bulgaria) and the Skinakas observatory of the Institute of Astronomy, University of Crete (Greece).  
The region of NGC 7000/IC 5070 is a large H II complex, one of the closest to our Sun and a very attractive field of intensive star formation in our Galaxy (Tsvetkov 1980).
The field is rich in young stars with different masses and it is the object of constant interest to researchers of star formation (Armond et al. 2011).
For several years we perform regular photometric observations in the field of the dark clouds ("Gulf of Mexico") in order to study the long-time variability of PMS stars.
Using our photometric observations from 2007, 2008, and 2009, a comparison sequence of fifteen stars in the field around HBC 722 was calibrated in
$BVRI$ bands (Semkov et al. 2010).  

The $\it BVRI$ light-curves of HBC 722 during the period April 2009 - August 2011 are plotted in Fig. 2.  
The photometric observation obtained before the outburst (Apr.  - Nov. 2009) displayed only small amplitude variations in all pass-bands.  
The observational data indicate that the outburst started sometime before May 2010, and reached its maximum value in September 2010.  
By comparing with brightness levels in 2009, we derive the following values for the outburst amplitude: $\Delta$$I$=3\fm7, $\Delta$$R$=4$\fm$3, $\Delta$$V$=4\fm7, and $\Delta$$B$=4\fm7.  
Simultaneously with the increase in brightness the star color changes significantly, becoming appreciably bluer.  
An important result from the $BVRI$ photometry is that, while both $V$$-$$R$ and $R$$-$$I$ became bluer, at the same time $B$$-$$V$ did not change relative to quiescence.  
Since October 2010, a slow fading was observed and up to May 2011 the star brightness decreased by 1\fm4 (V). 
In the period May - August 2011 no significant changes in the brightness of the star are observed, its brightness remains at 3\fm3 (V) above the quiescence level.
During the period of rise in brightness and the first months after the maximum, the light curve of HBC 722 is similar to the light curves of the classical FUor object V1057 Cyg and FU Ori itself (Clarke et al. 2005).

\begin{figure}[!htb]
  \begin{center}
   \centering{\epsfig{file=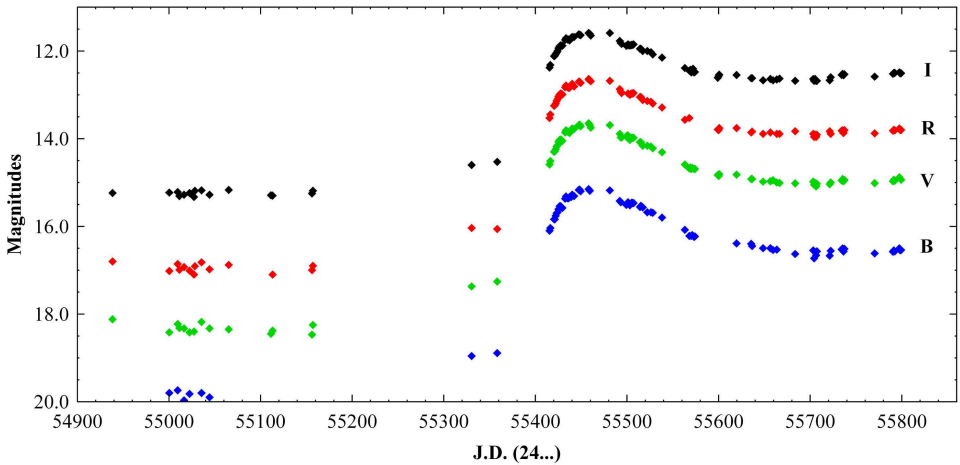}}
    \caption[]{$BVRI$ light curves of HBC 722 for the period Apr. 2009 - Aug. 2011}
    \label{countryshape}
  \end{center}
\end{figure}

Long-slit optical spectra of HBC 722 have been obtained with the 1.3-m RC telescope of the Skinakas Observatory on 2001 before the outburst and on 2010-2011 during the outburst.
In order to investigate the most prominent spectral futures for FUor stars the red spectral range (5500–7200 \AA) was observed.
Observations were carried out with ISA 612 spectral CCD camera ($2000\times800$ pixels, $15\times15$ $\mu$m scale) and a focal reducer. 
A reflection grating (1300 lines/mm) and 160 $\mu$m slit were used. 
This combination yields a resolving power $\lambda$/$\Delta\lambda$ $\sim$ 1300 at H$\alpha$. 

During the outburst the spectrum of HBC 722 changed significantly and from a typical spectrum of T Tauri star it became similar to spectra of other FUor objects. 
On the pre-outburst spectrum of HBC 722 obtained on September 2001 only two emission lines H$\alpha$ and [OI] 6300 are seen (Fig. 3).  
The broad emission hydrogen H$\alpha$ line is one of the most representative features of T Tauri stars. 
Since the spectra of Cohen \& Kuhi (1979) and our one in Fig. 3 are remarkably similar, we have reason to assume that HBC 722 has characteristics of a classical T Tauri star during the period before the outburst.
This statement is also supported by our photometric data obtained in the period before 2010.

\begin{figure}[!htb]
  \begin{center}
   \centering{\epsfig{file=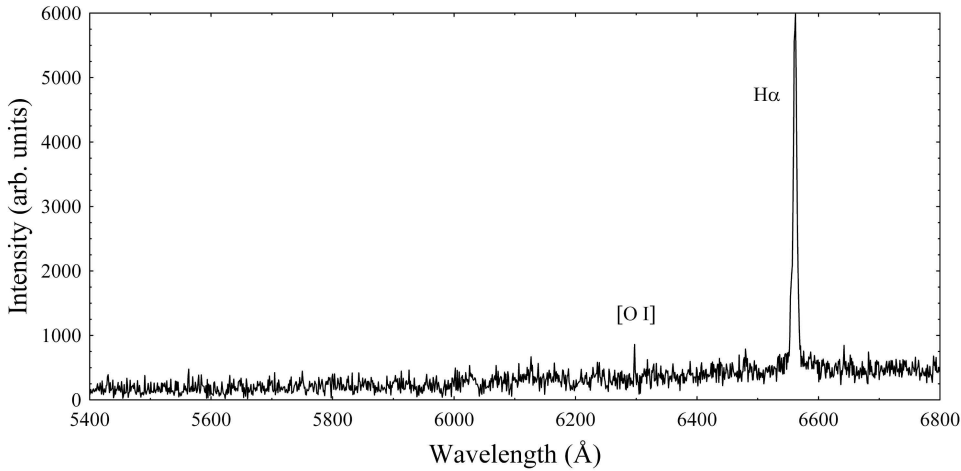}}
    \caption[]{Pre-outburst spectrum of HBC 722 obtained on 2001 Sep. 10}
    \label{countryshape}
  \end{center}
\end{figure}

\begin{figure}[!htb]
  \begin{center}
   \centering{\epsfig{file=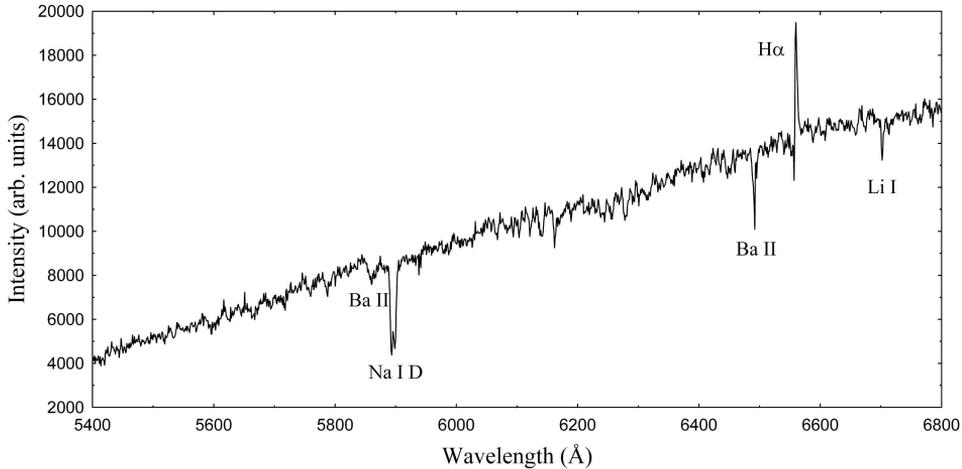}}
    \caption[]{Outburst spectrum of HBC 722 obtained on 2011 Aug. 15}
    \label{countryshape}
  \end{center}
\end{figure}

After the outburst the emission lines disappear from the spectrum of HBC 722. 
Only H$\alpha$ remains in emission, but it has a smaller equivalent width (Semkov et al. 2010).
The outburst spectrum is dominated by absorption lines of Na I ($\lambda$ 5890/5896), Li I $(\lambda$ 6707), and Ba II ($\lambda$ 5854, 6497) (Fig. 4).
As noted in Semkov et al. (2010) the profile of H$\alpha$ line is highly variable during the rise in brightness and maximum light, and the appearance of a strong absorption component is observed. 
The recent spectrum of HBC 722 shows a deepen absorption component of H$\alpha$ line shifted towards the blue spectral region.
Therefore, we observe the formation of the typical of FUor stars a P Cyg profile of H$\alpha$ emission line (Fig. 5).

\begin{figure}[!htb]
  \begin{center}
   \centering{\epsfig{file=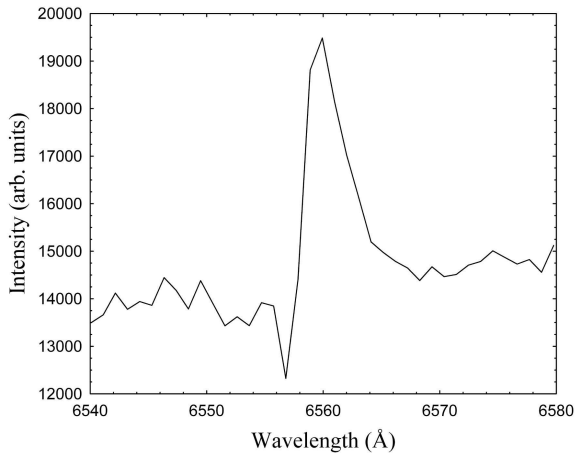}}
    \caption[]{The profile of H$\alpha$ emission line extracted from the spectra of HBC 722 obtained on 2011 Aug. 15}
    \label{countryshape}
  \end{center}
\end{figure}

\section{Observations of HBC 722 from observatories around the world}

During the year passing from the beginning of outburst the interest in HBC 722 continuously increase.
The first results from observations have been published in astronomical journals with high impact factor.
The outburst was independently discovered by Miller et al. (2011) during the regular monitoring of NGC 7000 with the Palomar 48-in telescope.
After the automatically reduction of obtained images a new variable sources (PTF 10qpf) were found using the software developed for the Palomar Transient Factory.
Authors confirm the FUor nature of HBC 722 eruption and note that it is the first FUor associated with a previously studied young stellar object.
Results from infrared photometry and spectroscopy, as well as low and high resolution optical spectroscopy (with the Keck I telescope and HIRES spectrometer) of the object are reported in the paper of Miller et al. (2011).
The results from high resolution optical spectroscopic monitoring campaign of HBC 722 during November - December 2010 are reported by Lee et al. (2011).  

The increase in the infrared brightness of HBC 722 was reported by Leoni et al. (2010), whose data for 2010 Sep. 1, when compared with 2MASS values provide the outburst amplitudes $\Delta$$J$=3$\fm$2, $\Delta$$H$=3\fm0, and $\Delta$$K$=2\fm8. 
The beginning of decrease in brightness in the near-infrared $JHK$ bands and in the optical $R$ band was reported by Lorenzetti et al. (2011).
$JHK$ near-infrared observations of HBC 722 obtained in autumn 2010 were also published in K{\'o}sp{\'a}l et al. (2011).

The pre-outburst spectral energy distribution of HBC 722 is discussed in the papers of Miller et al. (2011) and K{\'o}sp{\'a}l et al. (2011).
The authors concluded that before the eruption HBC 722 was a Class II young stellar object - most often associated with Classical T Tauri stars.
The calculated pre-outburst bolometric luminosity of the object is 0.85 $L_{\sun}$ (K{\'o}sp{\'a}l et al. 2011), while during the outburst it rise to  $\sim$12$L_{\sun}$ (Miller et al. 2011).
This value is at the bottom of the luminosity scale for FUor outbursts, but still comparable to the luminosity of some FUor objects like L1551 IRS5 and HH381 IRS (Reipurth \& Aspin 2010).

Green et al. (2011) analyze the submillimeter emission surrounding HBC 722 using images and spectroscopy from the Herschel Space Observatory and the Caltech Submillimeter Observatory. 
The authors detect CO emission in the surrounding region, evidence of outflow-driven heating in the vicinity of the object.
HBC 722 does not show evidence for a circumstellar envelope or shocked gas, and appears to have erupted from a disk-like state, similar to FU Orionis itself (Green et al. 2011).

\section{Conclusions}

Given the small number of known FUor objects, photometric and spectral studies of HBC 722 are of great interest.  
In the first year after the outburst results from following observations were published: optical photometry with 12 telescopes in 8 observatories, infrared photometry with 4 telescopes in 4 observatories, spectroscopy (optical and infrared) from low of high resolution with 8 telescopes in 7 observatories.
Results from three space missions - Hershel, Spitzer, and Swift were published also.

The photometric and spectroscopic data collected so far suggest that HBC 722 is indeed a FUor object.  
The $\Delta$V=4\fm7 amplitude, the spectrum turning from emission-line dominated to one with prominent absorption lines, and the emergence of a reflection nebula around the star are all well-established characteristics of FUor outbursts.  
The profile of the H$\alpha$ line is highly variable during the maximum light and the decrease in brightness, but a P Cyg profile characteristic of the FUor outburst is registered with confidence. 
We plan to continue with our spectroscopic and photometric monitoring as well as with a study of archival photographic observations.

{\it Acknowledgments:} This work was partly supported by grants DO 02-85, DO 02-273 and DO 02-362 of the National Science Fund of the Ministry of Education, Youth and Science, Bulgaria. 
This research has made use of the NASA Astrophysics Data System.

\end{document}